\documentclass[pra,twocolumn,showpacs,superscriptaddress,floatfix, nofootinbib]{revtex4-1}
\usepackage[caption=false]{subfig}
\usepackage{amsmath,graphicx}
\usepackage{amsmath,amssymb,mathrsfs,esint}
\usepackage{multirow}
\usepackage{algorithm}
\usepackage[noend]{algpseudocode}
\usepackage{comment}
\usepackage{tabularx}
\usepackage{bm}
\usepackage{braket}
\usepackage{enumerate}
\usepackage[normalem]{ulem}
\usepackage[bookmarks=true,
   colorlinks=true,
   linkcolor=blue,
   urlcolor=blue,
   citecolor=blue,
   bookmarks=true,
   hyperindex=true
]{hyperref}

\begin{document}

\title{Deploying hybrid quantum-secured infrastructure for applications: \\ When quantum and post-quantum can work together}

\author{Aleksey K. Fedorov}
\affiliation{Russian Quantum Center, Skolkovo, Moscow 121205, Russia}
\affiliation{National University of Science and Technology ``MISIS”, Moscow 119049, Russia} 

\begin{abstract}
The vast majority of currently used cryptographic tools for protecting data are based on certain computational assumptions, 
which makes them vulnerable with respect to technological and algorithmic developments, such as quantum computing. 
Existing options to counter this potential threat include, first of all, quantum key distribution, whose security is based on the laws of quantum physics.
Quantum key distribution is proven to be secure against any unforeseen technological developments.
The second approach is to switch to post-quantum cryptography, which is a set of cryptographic primitives that are believed to be secure even against attacks with both classical and quantum computing technologies. 
In this perspective, a recent progress in the deployment of the quantum-secured infrastructure based on quantum key distribution, post-quantum cryptography, and their combinations, is reviewed. 
Various directions in the further development of the full-stack quantum-secured infrastructure are indicated. 
Distributed applications, such as blockchains and distributed ledgers, are also covered. 
\end{abstract}

\maketitle

\section{Introduction} 

The number of data that human society create, collect, copy, transmit, and store increases extremely fast~\cite{Statista}.
Such digitalization increases demands for efficient methods for (long-term) protection of data.
Moreover, as it is estimated, the portion of data requiring protection also grows from year to year~\cite{Soller2020}. 

Among existing ways to protect the data in the current digital era, {\it cryptography} plays a central role in ensuring the security and privacy of information; its applications ranges from personal data to critical infrastructure.
A basic idea of cryptography is to realize {\it encryption}, i.e. a transformation of the data to the form, which makes it still accessible for legitimate sides of communications (commonly referred to as Alice and Bob), but not for a third unauthorized party (Eve). 
This allows one to prevent an unauthorized extraction of the information that is transmitted over an insecure channel~\cite{Schneier1996}.
An idea behind is that there is a certain parameter, which is known as a {\it cryptographic key}, determining the choice of a specific transformation of the information (among possible ones) when performing encryption.

As it was shown by Shannon~\cite{Shannon1948} and, independently, by Kotelnikov\footnote{V.A. Kotel'nikov, Classified Report (1941).}, 
there is a cryptographic algorithm, which is perfectly secure, namely the one-time pad, also known as the Vernam cipher~\cite{Vernam1926}. 
The perfect or information-theoretic security here means that the scheme remains secured even if one assume that the eavesdropper has unlimited computational resources (its abilities are limited only by laws of physics).
In this scheme a legitimate side of communication Alice encrypts her message, a string of bits denoted by the binary number $m$ with the use of a randomly generated key $k$ as $d=m\oplus{k}$, where $\oplus$ denotes modulo-2 summation.
The message is then sent to Bob, who decrypts the message as $d\oplus{k}=m\oplus{k}\oplus{k}=m$
However, the cost for such level of security is the fact that the used cryptographic key $k$ should be (i) random, (ii) use only once, and (iii) has the length in bits that is not less that the length of the message to be encrypted. 
A class of cryptosystems that use the same key for encrypting and decrypting data, as it takes place for the Vernam cipher, is known as {\it symmetric} cryptography (or private-key cryptography).

Cryptographic keys are then extremely valuable resource; as it is emphasized in Ref.~\cite{Schneier1996}:
``Keys are as valuable as all the messages they encrypt, since knowledge of the key gives knowledge of all the messages. 
For encryption systems that span the world, the key distribution problem can be a daunting task.'' 
Key distribution is then crucial for cryptography.
One of possible solutions is to use trusted couriers, which physically, by nondigital means transfer cryptographic keys from one place to another. 
Although such an approach may seem rudimentary, it is still used for specific applications~\cite{Mosca2017}. 
However, it is clear that in the era of digital communications is impossible to use trusted couriers to supply cryptographic keys to all the users of global, distributed communication networks.
Moreover, the human factor may impose additional threats:
Imagine that the courier known the key, then any data that is protected by means of this key is accessible for him unless the key is changed; 
moreover, it is almost impossible to determine the fact of the attack, so that this vulnerability can be used for sufficiently long period of time.
One can think of more frequent changes of the keys, which, however, increases the cost of the trusted-courier-based key distribution.
As it is also noted in the seminal paper by Diffie and Hellman~\cite{Diffie1976}, the cost and delay imposed by such a ``physical'' key distribution method is ``a major barrier''.
Existing symmetric cryptographic tools, such as AES (the Advanced Encryption Standard), deal with common private keys of lengths that are less than sizes of the messages. 
In this sense, the problem of their security belongs to the matter of practical (computational) security.

An alternative is to use so-called {\it public-key} or {\it asymmetric} encryption systems, 
which are based on the idea of reducing the problem of unauthorized access to information to solving a computational problem that is believed be hard~\cite{Diffie1976}. 
For example, multiplying two large prime numbers, $P$ and $Q$, is easy (at it is then easy to verify that multiplication of two prime numbers gives the correct integer number), 
but finding the prime factors of a given product $N$ is hard computational problem.
Under the assumption that existing computers could not solve such mathematical tasks in a reasonable time, modern public-key cryptography techniques, 
such as, for example, the Rivest-Shamir-Adleman (RSA) scheme~\cite{Rivest1978}, seem to be secure.
However, one must keep in mind that this an assumption, so it is not proven. 
Moreover, as it has been shown, quantum computing devices~\cite{Brassard1998,Ladd2010} are believed to be powerful in solving certain classically-difficult tasks, 
including prime factorization as it has been proposed by Shor~\cite{Shor1994}. 
Therefore, a paradigm of information protection in the era of quantum computing (``post-quantum era'') should be reconsidered:
At least one should keep in mind that an adversary can potentially use quantum computing devices to attack cryptographic tools.
This attack can be delayed in time, but still important for understanding the principles of information security in the post-quantum era.

The consequences of the appearance of a large-scale quantum computer that is able to attack, let say, RSA algorithm for realistic key sizes is then can be seen as a catastrophe
(here I refer the reader to the paper entitled as ``The Day the Cryptography Dies"~\cite{Mosca2017}). 
Fortunately, not all the tools are vulnerable to quantum cryptoanalysis.
Existing options to counter this potential threat include, first, quantum key distribution (QKD)~\cite{Gisin2001,Bennett1984,Ekert1991}.	
The idea behind QKD protocols is to use quantum objects instead of physical couriers for cryptographic keys.
The security proof of QKD~\cite{Gisin2001,Scarani2009,Gisin2014,Kiktenko2016,Renner2022} is based on the laws of quantum physics and, 
thus, it is guaranteed to be secure against any unforeseen technological developments, including, for example, even more powerful quantum computers.
A class of attacks on QKD systems may be split between two major issues: the first is to prove that the protocol is indeed secure and the second is to illuminate technological drawbacks in implementing QKD devices. 
The second approach is to switch to post-quantum cryptography~\cite{Bernstein2017}, which is based on cryptographic primitives that believed to be secure even against attacks with quantum technologies. 
Mathematical tools of post-quantum cryptography include hash-based, lattice-based, code-based, and other approaches having their own advantages and disadvantages (for a review, see Ref.~\cite{Bernstein2017}).
Both these methods --- QKD and post-quantum cryptography --- have their own prospects and limitations, which are needed to be taken into account when deploying them.

In this perspective, I review a recent progress in the deployment of the quantum-secured infrastructure based on quantum key distribution, post-quantum cryptography, and their combinations. 
We indicate various direction in the further development of the full-stack quantum-secured infrastructure and argue that the hybrid approach, which combines quantum key distribution and post-quantum cryptography, can be beneficial for various applications. 
We specifically focus on distributed applications, such as blockchains and distributed ledgers. 

This perspective is organized as follows.
In Sec.~\ref{Sec:QSC}, we review quantum-secured cryptography tools with highlighting their advantages and limitations.
In Sec.~\ref{Sec:Hybrid}, we discuss hybrid approaches, in which quantum key distribution works jointly with post-quantum primitives.
We conclude in Sec.~\ref{Sec:Outlook}.

\section{Quantum-secured cryptography}\label{Sec:QSC}

\subsection{Quantum cryptoanalysis}

New approaches for designing computational devices may lead to the need in modifying assumptions about the security of certain cryptographic primitives.
A celebrated example is that Shor's algorithm~\cite{Shor1994,Shor1999} for solving the prime factorization and discrete logarithm problems in polynomial time. 
Proof-of-concept experimental factoring of 15, 21, and 35 have been demonstrated on superconducting~\cite{Martinis2012-2}, trapped ion~\cite{Blatt2016}, and photonic~\cite{Pan2007,White2007,OBrien2012} quantum computers. 
Generally, a variant of Shor’s algorithm by Beauregard in Ref.~\cite{Beauregard2003} requires $\mathcal{O}(n^3\log{n})$ operations under $2n+3$ logical (ideal) qubits if $N={p}\times{q}$ fits into $n$ bits~\cite{Bernstein2017}.
Having in mind the need in quantum error correction, 
Shor’s algorithm for practically relevant key sizes (for example, 2048 bit) would need 8 hours using 20 million physical qubits~\cite{Gidney2021}, which greatly exceeds the capabilities of today's quantum computing devices. 
Another recent proposal~\cite{Gouzien2021} demonstrates a way to factor 2048 RSA integers in 177 days with 13436 physical qubits and a multimode memory. 
A forecast review~\cite{Sevilla2020} estimates the likelihood for quantum devices capable of factoring RSA-2048 to exist before 2039 as less than 5$\%$. 

Quantum computing also has an impact on symmetric cryptography, such as AES encryption.
Grover's algorithm~\cite{Grover1996} enables quadratic speedup in brute force search, which means that the key length should be doubled to enable the same level of protection~\cite{Kim2018}. 
The same scaling applies to cryptographic hash functions, for which the primary attack method is also brute-force search~\cite{Kim2018}. 
In certain cases, this could be offset by increasing the length of the symmetric key. 

Although a problem of breaking certain cryptographic tools may seem abstract and far from end users, there are several arguments to take such a risk into account.
The first argument is related to the so-called ``store now --- decrypt later`` attack~\cite{Mosca2017}. 
The idea is that the adversary is harvesting information in the encrypted form, in the hope that new computational devices, for example, quantum computing, will help them to uncover valuable information from it in the future. 
That is why for some particular applications dealing with long-term sensitive information (such as medical records or genetic data), one should think about the priority replacement of cryptographic primitives. 
This fact is expressed in Mosca’s theorem~\cite{Mosca2017,Mosca2018}: 
We need to start worrying about the impact of quantum computers when the amount of time that we wish our data to be secure for ($X$) added to the time 
it will take for our computer systems to transition from classical to secured against quantum attacks ($Y$) 
is greater than the time it will take for quantum computers to start breaking existing quantum-susceptible encryption protocols ($Z$).

The second argument that the security problems impacts not only on encryption, but also on digital signatures, which are widely deployed across governmental services and business applications:
One needs to replace the whole system of digital signatures in order to use them in the quantum computing era.
Therefore, for each information system one need to formulate the transition plan and formulate security recommendations for public-key infrastructures in the post-quantum era
(see Ref.~\cite{Yunakovsky2021} as an example for the case of recommendations for production environments).

Finally, it is extremely hard to predict the time of the appearance of a large-scale quantum computing device: 
An expected breakthrough in quantum computing, both in hardware and algorithms, is possible, may quickly change the situation and actualizes this problem. 
In particular, no fundamental obstacles preventing quantum computer from further scaling have been identified up to date. 
As an example, there is an increasing interest in alternative schemes for solving the prime factorization problem using quantum tools, such as variational quantum factoring~\cite{Karamlou2021,Aspuru-Guzik2019}.
A very recent proposal demonstrates a possibility to accelerate factoring using quantum computers with sublinear scaling in the number of qubits~\cite{Yan2022} (we, however, warn the reader since this proposal is not yet fully verified).
To summarize, we are in a race against time to deploy quantum-safe cryptographic tools, which are protected both from attacks with classical and quantum computers, before powerful enough quantum computing devices appear.

\subsection{Quantum key distribution}

What is the basic problem of symmetric cryptographic primitives? 
It is to ensure that the key distribution processes is organized properly. 
As we have discussed before, the cost of the key distribution process is extremely high, especially in the case of usage one-time pad encryption. 

A beautiful idea is to replace vulnerable cryptographic primitives with one-time-pad encryption with QKD~\cite{Bennett1984,Ekert1991},
which is a technology of using individual quantum objects to establish cryptographic keys. 
Historically, one of the first BB84 QKD protocol~\cite{Bennett1984} has developed the idea of conjugate coding~\cite{Wiesner1983} without using the feature of quantum entanglement,
whereas the protocol proposed independently by Ekert~\cite{Ekert1991} uses so-called entangled quantum states.
Information carriers in QKD systems are photons since they perfectly suit for this task both in classical and quantum domains (QKD is able to use fiber-based and free-space communication channels). 
The fundamental advantage of this approach is that the QKD security relies not on any computational assumptions, 
but the laws of quantum physics~\cite{Gisin2001,Mayers2001,Shor2000,Koashi2009,Scarani2009,Tomamichel2012,Renner2022}. 
Below we stress on BB84 QKD protocol.

The concept of QKD is that two legitimate users (Alice and Bob) have {\it the pre-shared authentication key} (this aspect is discussed below) and the communication channel. 
They use a certain protocol for preparing quantum states and encoding information on the Alice side, and measure the states on the Bob side. 
Here and below we follow Ref.~\cite{Trushechkin2021} in explaining the basics of QKD.
Alice and Bob use four qubit states, which form two orthogonal bases $z{=}\{\ket0_z,\ket1_z\}$ and $x{=}\{\ket0_x,\ket1_x\}$ in the two-dimensional Hilbert space.
The values 0 or 1 indicate, which classic bit is encoded by the corresponding basis vector. 
Elements of the bases are expressed in terms of elements of another basis according to the relations
\begin{equation}\label{eq:bb84basis2}
	\ket0_x=\frac{\ket0_z+\ket1_z}{\sqrt2},
	\qquad
	\ket1_x=\frac{\ket0_z-\ket1_z}{\sqrt2}.
\end{equation}

If the information is encoded into photon polarization, then the vectors $\ket0_z$ and $\ket1_z$  can correspond, for example, to horizontal and vertical polarizations. 
In this case, $\ket0_x$ and $\ket1_x$ correspond to two diagonal polarizations that are rotated by 45$^\circ$ and 135$^\circ$ degrees, respectively, relative to the horizontal direction. 
The polarization coding is used to illustrate the idea, but, in fact, there is no restriction on the method of information encoding: 
Formally, $\ket0_z$, $\ket1_z$, $\ket0_x$, and $\ket1_x$ are vectors in the Hilbert space and one can use any encoding which fulfills relation given by Eq.~(\ref{eq:bb84basis2}).
The equivalence of two populars ways of encoding --- polarization and phase encodings --- is in detail explained in Ref.~\cite{Trushechkin2021}.

Importantly, as is seen from (\ref{eq:bb84basis2}), when measuring a qubit in a basis different from the preparation basis, the result is a random value. 
In opposite, if bases of preparation and measurement coincidence, 
the result perfectly correlates with the prepared state of the qubit 
(in the absence of errors in the channel, measuring devices and so on).

The BB84 protocol works as follows~\cite{Bennett1984,Gisin2001,Trushechkin2021}:
\begin{enumerate}
	\item 
	Alice randomly chooses a basis from the set $\{z,x\}$ and the value of the transmitted bit of information: $1$ or $0$. 
	Bits are selected with equal probabilities of 1/2.
	\item
	Then, the photons prepared in the corresponding states are transmitted through the communication channel.
	\item
	Bob randomly chooses a measurement basis, $z$ or $x$, for each qubit and measures the state of the qubit in the selected basis.
	If the preparation and measurement bases coincide, the received bit value coincides (ideally) with the sent one.
	If the bases do not coincide, the bits of Alice and Bob do not correlate (that is, they may or may not coincide with equal probabilities) due to the fact that the bases are mutually unbiased~(\ref{eq:bb84basis2}). 
	Usually, the communication channel contains large losses; 
	therefore, not all positions are registered by the receiver.
	\item
	The above steps are repeated many times, i.e., a large number of quantum states are transmitted. 
	As a result, legitimate parties receive two sequences of bits $k^{\rm raw}_A$ and $k^{\rm raw}_B$, which are called the {\it raw quantum-generated keys}.
\end{enumerate}
Since a perfect copy of a quantum state cannot be created~\cite{Dieks1982,Wootters1982} 
and the adversary does not know the basis, in which the bit is encoded in a given position, the adversary needs to employ imperfect copying techniques that induce errors.
We note that there are also so-called continuous-variables QKD protocols, which we do not cover in this paper (see Refs.~\cite{Gyongyosi2020,Pirandola2020}).

At the second stage, Alice and Bob use the classical post-processing of raw quantum-generated keys, $k^{\rm raw}_A$ and $k^{\rm raw}_B$, with the use of 
communications a over public authenticated channel~\cite{Gisin2001,Ma2010,Gisin2014,Kiktenko2016}:
\begin{enumerate}
	\item 
	{\it Announcements}. 
	Bob announces the position numbers, in which the signal has been registered.
	Alice and Bob then discuss the bases used in all positions.
	When using the decoy state method (see below), Alice also announces the type of each pulse (signal or decoy) at this state.
	Alice and Bob can also announce bits in positions that do not participate in the formation of the secret key: 
	In positions in which the parties used the $x$ basis and in the decoy pulses.

	\item
	{\it Key sifting}.
	Positions in which the decoy state intensity has been used, registration did not occur, or at least one of the legitimate parties used the $x$ basis are sifted out.
	The resulting keys, $k^{\rm sift}_A$ and $k^{\rm sift}_B$, are called {\it sifted keys}.
	Ideally, they should match, but as a result of natural noise in the channel or adversary actions, 
	they do not match.
	Moreover, the adversary (Eve) may have partial information about them.
	
	\item
	{\it Error correction}.
	One of the sifted keys (for example, belonging to Alice) is considered as a reference. 
	Differences between it and the sifted key of the other side are considered to be caused by errors.
	One can use error correction codes or interactive error correction procedures to correct errors.
	Low-density parity-check (LDPC) codes are commonly used for this purpose.
	Often, this procedure ends with {\it verification}: 
	The identity of the sifted keys is checked using hash functions (see Ref~\cite{Fedorov20182}).
	As a result of this stage, the legitimate parties receive identical {\it verified keys} 
	$k^{\rm ver}_A=k^{\rm ver}_B$ with a high probability.
	An efficient method for error correction in the BB84 protocol error correction based on LDPC codes is described in Ref.~\cite{Kiktenko2017}; 
	see also Ref.~\cite{Kronberg2021} for progress in LDPC codes and Ref.~\cite{Kiktenko2020} for polar codes.

	\item
	{\it Estimation of the level of eavesdropping} and making a decision about creating the key or renouncing it (aborting the protocol) based on the observed data.
	QKD protocols are based on the fact that information encoded in non-orthogonal quantum states cannot be read by a third party 
	(which does not know the basis in which the key bit in a given position was encoded) 
	without ``spoiling'' these states.
	Therefore, any interception by Eve would lead to an increase in the number of errors (i.e., mismatched positions in sifted keys) between the legitimate parties.
	In this version of the protocol, where only the bits encoded in the $z$ basis are involved in the formation of the key, 
	only the fraction of errors in the $x$ basis is needed to assess the level of eavesdropping.
	If the error rate exceeds a certain critical threshold, the protocol is aborted with a warning message.
	Otherwise, the parties proceed to the last step.
	The basic idea behind the implementation of QKD protocols is to create conditions, in which eavesdropping it is impossible without being detected.
	
	\item 
	{\it Privacy amplification.} 
	Alice randomly chooses a so-called hash function from a family of $2$-universal hash functions and sends it to Bob over a public channel.
	Then they compute the hash value of their (identical) sifted keys.
	As a result, Alice and Bob obtain a common shorter key (\textit {final key}) $k^{\rm fin}_A=k^{\rm fin}_B$, but the information of the adversary about which is now negligible.
	With an infinitely large length of the sifted key, it can be made arbitrarily small.
	The more information the adversary has about the sifted key (as a result of eavesdropping and as a result of disclosure by legitimate users of some of the information during error correction),
	the more compression of the key in the privacy amplification procedure is required, 
	i.e., the shorter the final key and the lower the key rate.
	
\end{enumerate}

The final key is split into two parts: the first is used for authentication for the next QKD rounds (as soon as this portion is needed to be minimized, the problem of optimizing the resources for QKD authentication appears, see Ref.~\cite{Kiktenko2020})
and the rest can be used for external applications.
By now the seminal BB84 protocol for QKD~\cite{Bennett1984,Gisin2001,Trushechkin2021}, which is typically accomplished by the decoy-state method~\cite{Trushechkin2021}
(in order to avoid the photon-number splitting attack, which appears due to the fact that Alice use weak laser pulses instead a true single photon source~\cite{Brassard2000}), 
is by now considered as a candidate for the standard\footnote{K. Chen, J. Ma, and H. Shi, Talk, ISO/IEC JTC1 SC27 WG3 SP Proposal, 
Security Requirements, Test and Evaluation Methods for the Decoy State BB84 Quantum Key Distribution (QKD), Berlin, Germany, 10/31/2017; 
ISO/IEC JTC 1/SC 27/WG 3 N 1537, 30th ISO/IEC JTC1/SC27 Working Group Meeting, H. Shi, J. Ma, and G. Pradel Wuhan, China, April 2018 30th Security Requirements, Test and Evaluation Methods for Quantum Key Distribution}.

As a result of the QKD session, Alice and Bob have a key for external applications, such as one-time pad encryption or AES block ciphers, and used to frequently refresh keys~\cite{Gisin2001}.
Such quantum-generated keys are proven to be information-theoretically secure against arbitrary attacks, including the quantum ones. 
Recent progress in development and commercialization of QKD systems is a significant step towards to improve secrecy of information. 
However, still there are several challenges on the way to wider adoption QKD technology. 

\subsubsection*{Practical aspects}

The first aspect is problem of security of QKD protocols. 
Although, it is largely accepted that the decoy-state BB84 protocol has its verified security proofs
(one of the latest developments here is related to security of quantum key distribution with detection-efficiency mismatch in single-photon~\cite{Trushechkin2019} and multiphoton~\cite{Lutkenhaus2021,Trushechkin2022} cases), 
various many alternatives to this protocol have been proposed~\cite{Pirandola2020}.
These alternative may seem interesting for achieving higher key generation rates or simpler practical implementation, 
however, their security proofs require further analysis --- in order to achieve at least such level as in the case of the decoy-state BB84 protocol~\cite{Pirandola2020,Trushechkin2021}.
That is why currently availably QKD implementations are mostly based on the decoy-state BB84 protocol.
Thus, any alternative protocols should be in detail investigated before being used in industrial QKD systems.

The second question is related to the efficiency of the post-processing procedure.
A remarkable progress over last decade in making all the steps of the procedure has been performed.
However, there is still a room improvement. 
A specific direction is the optimization of post processing with respect the the network topology.
For example, Ref.~\cite{Tayduganov2022} proposes the way to asymmetric error correction that could be used in practical QKD systems with limited computational resources on of the sides.
Also, new types of error correction codes, such as polar codes, can be considered~\cite{Kiktenko2020}.

The third aspect is the distance issues.
An efficient performance of QKD devices over long distances ($\geq$500 kilometers) remains an serious challenge due to optical losses over the entire communication distance~\cite{Muralidharan2016}
(we note that there are sever remarkable experiments on long-distance QKD with the twin-field protocol~\cite{Wang20222}).
The range of commercial QKD systems is typically 100 kilometers over optical fibers. 
Despite the tremendous work on creating new protocols~\cite{Wang20222}, quantum repeaters~\cite{Muralidharan2016,Lukin2020-10} and using satellites for global-scale QKD~\cite{Pan2017} (for a review, see Refs.~\cite{Bedington2017,Pan2022}), 
QKD technology still faces several challenges~\cite{Lo2014,Lo2016}, 
which makes it best suitable for some domain-specific applications, such as the protection of highly-loaded communications links at a distance, which does not require the use of intermediate nodes. 
As it has been discussed, long-range QKD without trusted nodes is not possible with current technology~\cite{Zbinden2022}.
Therefore, an important question is to optimize the performance of large-scale backbone QKD networks. 
One of the ideas for this purpose is to use switch-based QKD backbone networks with trusted repeaters
As it is estimated in Ref.~\cite{Tayduganov2021} 
for a network link of a total of 670 km length consisting of 8 nodes and demonstrate that the switch-based architecture achieves significant resource savings of up to 28\%, while the throughput is reduced by 8\% only.

Fourth question is about practical security of QKD devices, i.e. a class of attacks that appear due to the fact that real QKD devices do not exactly follow the underlying theoretical models (in particular, because of various engineering issues). 
We would like to acknowledge intensive research on finding various practical imperfections in QKD systems (for example, see Ref.~\cite{Makarov2011}).
This ``quantum white-hat hacking'' activity is highly important towards further process of QKD devices.

Finally, one needs to have quantum-secured cryptographic primitives for various problems beyond the key distribution problem~\cite{Broadbent2016}.
For example, we need digital signatures, which in principle can be realized using QKD~\cite{Gottesman2001,Andersson2016,Kiktenko2022}, but it is practically challenging to deploy the corresponding infrastructure. 

\begin{table*}[]
\begin{tabular}{|l|l|l|l|}
\hline
\textbf{Cryptographic algorithm} & \textbf{Type} & \textbf{Purpose}            & \textbf{Quantum security} \\ \hline
AES                              & Symmetric     & Encryption                  & Larger key sizes needed   \\ \hline
SHA-2, SHA-3                     & --            & Hash functions              & Larger output needed      \\ \hline
RSA                              & Public key    & Signature, key distribution & No longer secure          \\ \hline
ECDSA, ECDH                      & Public key    & Signature, key distribution & No longer secure          \\ \hline
DSA                              & Public key    & Signature, key distribution & No longer secure          \\ \hline
\end{tabular}
\caption{Security of cryptographic algorithms in the post-quantum era, see Ref.~\cite{Fedorov2022}; see also Ref.~\cite{Bernstein2017}.}
\label{tab:security}
\end{table*}

\subsection{Post-quantum cryptography} 

Fortunately, there are several tools helping to provide security even under the assumption that the eavesdropper has a large-scale quantum computer~\cite{Bernstein2017}.
One of the ideas is to use another class of computational problems, which are not vulnerable to attacks with quantum computers (see Table~\ref{tab:security}).
As it is mentioned above, quantum computing also has an impact on symmetric cryptography since quantum Grover’s algorithm provides a quadratic speed-up in the brute force search.
However, quadratic speed-up seem to be not dramatic since doubling the key size helps to eliminate this effect. 
At the same time, if the key is distributed by means of non-quantum-secured tools, the system would not guarantee security in the post-quantum era. 
Finding the ways to ensure quantum-secured key generation processes is crucial for various applications, in particular, Transport Layer Security (TLS), which is the security protocol behind the Hypertext Transfer Protocol Secure (HTTPS).
Therefore, the question is how to design quantum-secured for key distribution and digital signatures.
Several cryptosystems for these purposes, which strive to remain secure under the assumption that the attacker has a large-scale quantum computer, have been proposed.
Such an approach is known as {\it post-quantum cryptography}.
Its main advantage is the ability of relatively cheap and fast switching to new post-quantum algorithms (the $Y$ is minimized in terms of the Mosca’s theorem).

Post-quantum protocols are based on different mathematical approaches, such as 
(i) the shortest vector problem in a lattice~\cite{Micciancio2002,Hanrot2007,Regev2009}, 
(ii) learning with errors~\cite{Regev2010, Albrecht2015, Kirchner2015, Arora2011, Schnorr1994,Chen2011}, 
(iii) solving systems of multivariate quadratic equations over finite fields~\cite{Patarin1996,Faugere2003,Beullens2017}, 
(iv) finding isogenies between elliptic curves~\cite{Jao2011, Costello2017, Costello2016, Koziel2017, Steven1999, Delfs2016, Zhang2005, Tani2007}, 
(v) decoding problems in an error-correcting code~\cite{Berlekamp1978, Alekhnovich2003, May2015, Becker2012, Bernstein2010},
(vi) security properties of cryptographic hash-functions~\cite{Buchmann2011, Hulsing2016, Bernstein2019}, and other primitives~\cite{Bernstein2017}.
(see also NIST website\footnote{https://csrc.nist.gov/pro jects/post-quantum-cryptography} for existing submissions and Refs.~\cite{Bernstein2017,Yunakovsky2021} for references).
In this work, we do not provide detailed description of these primitives, which can be found in Refs.~\cite{Bernstein2017,Yunakovsky2021}.

\subsubsection*{Practical aspects}

Developing reliable security analysis for post-quantum algorithms is challenging.
A number of post-quantum cryptographic systems, which use post-quantum methods, 
are considered as candidates in the National Institute of Standards and Technology (NIST) Post-Quantum Cryptography Standardization and by European Telecommunications Standards Institute (ETSI).
Standardization processes are crucial for reducing cryptographic risks.
There are known examples of finding possible classical attacks to a post-quantum algorithm even at the mature stage of the standardization procedure.
New details in the security proofs also appear (for example, see Ref.~\cite{Kudinov2022} as the reply to Ref.~\cite{Kudinov2021}).
Another recent example is that an efficient key recovery attack on the Supersingular Isogeny Diffie-Hellman (SIDH) protocol has been proposed~\cite{Castryck2022}.
Moreover, one should keep in mind, again, that the assumptions about post-quantum algorithms can be reduced to statements based on computational assumptions. 
In a sense, the status of post-quantum cryptography is equivalent to the status of currently deployed public-key algorithms under the assumption of the absence of quantum computers.

An interesting aspect is that certain primitives, if used in specific protocols, have a feature of a possibility to detect their hacking, as it has been shown in the case of hash-based digital signatures~\cite{Kiktenko20212}.
In Ref.~\cite{Kiktenko20212} it has been shown that with properly adjusted parameters Lamport and Winternitz one-time signatures schemes could exhibit a forgery detection availability property. 
This property is of significant importance in the framework of the crypto-agility paradigm since the considered forgery detection works 
as an alarm that the employed cryptographic hash function becomes insecure to use and the corresponding scheme has to be replaced.

We also would like to note that the mitigation to post-quantum cryptography infrastructure for realistic cases requires deep analysis of related aspects, 
such as universal security requirements for the used software and SDKs especially in terms of the update policy.
A detailed description of security recommendation for public key infrastructures (PKIs), which are used as a part of security systems for protecting production environments, is presented in Ref.~\cite{Yunakovsky2021}.
Finally, it is widely discussed that post-quantum algorithms can be more resource-demanding rather than existing tools. 
Time consumption and memory consumption of post-quantum digital signature schemes is also presented in Ref.~\cite{Yunakovsky2021}.

\section{Inherently hybrid approach}\label{Sec:Hybrid}

A more close look at the QKD systems shows that such systems are inherently hybrid in a sense that quantum tools are essentially {\it combined} with the classical cryptography~\cite{Fedorov2018-3}. 
It is important to note that quantumness plays an important role at the stage of solving the key distribution problem, but not typically at the level of data protection.

\subsection{A peer-to-peer topology}

\subsubsection{Authentication}

The initial state of realizing QKD protocols requires authentication between the users~\cite{Gisin2001,Ma2010,Gisin2014,Kiktenko2016}. 
Typically, a classical approach of pre-shared cryptographic keys jointly with the Wegman-Carter scheme~\cite{Wegman1981}, which provides information-theoretic security. 
From this perspective, the QKD workflow appears to be {\it a key growing process}, since the parties already need to have a short pair of pre-distributed keys before the launching the first QKD round. 
For authentication in the second and subsequent rounds, the parts of quantum-generated secret keys from the previous round could be used~\cite{Kiktenko2020}. 

A possible approach to solve the authentication problem for large-scale QKD systems is to use post-quantum security for solving the problem of authentication.
As it noted in Ref.~\cite{Lutkenhaus2010}: ``If authentication is unbroken during the first round of QKD, even if it is only computationally secure, then subsequent rounds of QKD will be information-theoretically secure''
(for a review, see also Ref.~\cite{Allaume2014}).
Specifically, it has been shown that post-quantum digital signature can be used for authentication purposes in QKD devices~\cite{Yong2021,Wang2021}.
An important aspect is that one need to assume only the short-term security of post-quantum cryptography algorithms to achieve long-term security of the distributed keys,
since on next rounds one can also use quantum-generated keys for authentication purposes or their mixture. 
This systems is also of interest for QKD networks (see below) since for a QKD network of $n$ nodes in the case of using pre-sharing symmetric keys, 
$n(n-1)/2$ pairs of symmetric keys to realize pairwise interconnection are required. 

\subsubsection{Hybrid QKD protocols}

An original version of the BB84 QKD protocol assumes the choice of bases $\{z,x\}$ with equal probabilities of $1/2$.
Later, an improved variant of the protocol has been proposed, in which one of the bases 
(for example, the $z$ basis) is chosen more frequently than the other one~\cite{Lo2005}. 
This reduces the number of basis mismatches and, therefore, 
the portion of sifted positions, i.e., it increases the quantum-generated key rate.

This idea can be extended to a QKD protocol, in which the bases are chosen pseudo-randomly 
using a pre-distributed random sequence (probably, an portion of the authentication key). 
Such a modification of the BB84 protocol is considered in Ref~\cite{Trushechkin2018}.
As it has been shown, for single-photon sources, the considered protocol gives better secret key rates than the BB84 and the asymmetric BB84 protocols. 
However, the protocol strongly requires single-photon sources.

\subsubsection{Hybrid encryption}

Quantum-generated keys are then used for encryption, which is again purely based on the principles of classical cryptography. 
The combination of QKD with the one-time pad encryption would enable an information-theoretic secure cryptographic scheme~\cite{Gisin2001}. 

However, most of industrially-available encryptors use standartized symmetric protocols, such as AES, which is post-quantum under the assumption that the key has been generated via a quantum-secured approach~\cite{Bonnetain2019}. 
In order to make keep its security level in era of quantum computing, doubling key size is also a necessary step (this is due to the quadratic speed-up in the brute-force search provided by Grover's algorithm).
In this context, one of the scenarios related to the practical use of QKD in cryptographic infrastructures is to employ QKD as a key renewal technique for a symmetric cipher, such as AES, over a point-to-point link~\cite{Allaume2014}.
Also, symmetric quantum-generated keys can be efficiently combined with asymmetric post-quantum keys in various security models
(for example, see Ref.~\cite{Bogomolec2019} for the hybrid encryption scheme with session and public keys).

\subsection{Network applications for multiple users}

A natural application of QKD for multiple users is to consider a network containing many users with the objective of offering any-to-any key establishment service~\cite{Allaume2014}.
Such a scheme is easy to realize for networks with all-to-all topology, which is however challenging from practical point of view.
Long-range QKD without trusted nodes is not practically realizable with current level of technology~\cite{Zbinden2022}.
A hybrid quantum-secured infrastructure may use QKD for protecting highly-loaded communications link at the distance, 
which do not require the use of intermediate nodes, whereas end-users without direction connections can be protected by means of post-quantum cryptography.
Another schemes with hybrid (QKD with post-quantum) security can be also considered. 

\subsubsection{Post-quantum protection of trusted nodes}

One of the issues of implementing QKD protocols is the limitation related to the distance (see discussion above).
As it is mentioned above, long-range QKD without trusted nodes is not possible with current technology~\cite{Zbinden2022}.
It seems to be reasonable to have additional post-quantum authentication of the trusted nodes. 
This can be especially important for QKD networks without fully connected topology. 

\subsubsection{Quantum-secured blockchains and blind computing}

An area of particular concern in the context of quantum security is blockchains and cryptocurrencies~\cite{Fedorov2018,Fedorov2018-2,Tomamichel2018}.
Typical blockchain and cryptocurrency protocols use several cryptographic tools.
First, blockchains use as digital signatures to confirm the authorship of transactions.
Second, hash functions are employed for achieving a consensus (proof-of-work) between users in the absence of trust. 

Digital signatures that typically used in blockchains are based on primitives that are vulnerable to attacks with quantum computers.
For example, Bitcoin uses  the elliptic curve signature scheme.
As it is predicted in Ref.~\cite{Tomamichel2018}, by the most optimistic estimates such schemes could be completely broken by a quantum computer as early as 2027.
One of the potential applications of quantum computers on the bright side of cryptoanalysis can be finding lost private keys of legitimate users of cryptocurrencies. 
The quantum vulnerability of hash functions is similar to that of AES since the attack is based on brute-force search~\cite{Kim2018}, which can be enhanced by Grover algorithm in the quantum domain. 
Again, as it is discussed Ref.~\cite{Tomamichel2018}, blockchains that use proof-of-work consensus mechanisms, such as Bitcoin, are relatively resistant in the near-term horizon. 
We should mention that various blockchain platforms, such as Ethereum, mitigate to proof-of-stake consensus\footnote{https://ethereum.org/en/developers/docs/consensus-mechanisms/pos/},
which can make them more resistant to aforementioned types of attacks. 

Attacks with quantum computers have become a subject of many studies that proposed solutions for quantum-resistant blockchains~\cite{Fedorov2018,Fedorov2018-2}: 
blockchains that use quantum key distribution~\cite{Gisin2001} or post-quantum digital signatures and consensus schemes. 
A quantum-secured blockchain protocol was experimentally demonstrated in 2018~\cite{Fedorov2018}. 
There are several proposals how to realize quantum-secured blockchains with the use of entangled states~\cite{Farouk2017,Krishnaswamy2020,Nimbe2022}, 
which generally follow the initial idea of quantum solution to the Byzantine Agreement Problem~\cite{Gisin20012}.
There are several options to combine QKD with post-quantum security in blockchain networks~\cite{Fedorov2018-2}.
This is indeed of importance since the primary requirement to the original quantum-secured blockchain protocol is to have all-to-all connected QKD network, in order to implement the broadcast protocol. 

In addition to blockchains, another relevant application of both QKD and post-quantum cryptography is secure remote / blind computing, which is especially actual in the quantum domain~\cite{Broadbent2009}.
The idea behind is to ensure that a remote user can delegate a computational problem with a desire to keep the computation perfectly secret from untrusted servers implementing the quantum computation. 
Various cryptographic protocols for blind quantum computing have been proposed and tested~\cite{Aharonov20082,Dunjko2012,Morimae2012,Morimae2013,Mantri2013,Reichardt2013,Barz2013,Fisher2014,Morimae2014,Hayashi2015,Gheorghiu2015,Greganti2016,Marshall2016,Gheorghiu2017,Fitzsimons20172,Ma2022} (for a review, see Ref.~\cite{Fitzsimons2017}).
A proof-of-principle realization of blind quantum computing for completely classical clients has been recently presented~\cite{Pan2017-2}.
Further developments in this domain are required since both classical communication channels and directly themselves computational algorithms should be protected keeping in the mind the treat coming from quantum computers.

\section{Discussion and outlook}\label{Sec:Outlook}

An expected breakthrough in quantum computing, which possesses a significant threat on the currently widely deployed techniques for encrypting and protecting data, actualizes the problem of protecting data.
This is because most of cryptographic tools, which are difficult or impossible to break using conventional computing, become fairly easy to destroy using large-scale quantum computing devices.

The effect of quantum computing on information security can be mitigated by upgrading information security protocols with the use of QKD networks or post-quantum technologies. 
Thus, we are in a race against time to deploy quantum-safe cryptography, that it protected both from attacks with classical and quantum computers, before powerful enough quantum computers arrive~\cite{Mosca2017,Fedorov2018,Wallden2019}. 
For example, a present-day hacker might intercept and store encrypted messages with the hope to decrypt them with a quantum computer a few years later. 
If the information is long-term sensitive (medical records, genetic data, strategic plans, etc.), this attack may result in damages.  
In this domain, we see various opportunities for combining QKD security with its physics-laws-based security level and post-quantum cryptography that is able to substantially enhance the security level of information exchange protocols.
We also would like to note that at the current technology level QKD cannot cover all the required cryptographic primitives~\cite{Broadbent2016}, 
so there are plenty of aspects, in which quantum and post-quantum approaches can efficiently work together. 

{\bf Acknowledgements}.
The author thanks E. Kiktenko for useful comments and A. Mastiukova for inspiring discussions as well as the help with preparing the manuscript. 
We also thank the anonymous referee for useful comments, especially, pointing out secure remote computing as a potential application for quantum-secured tools. 
The analysis of quantum information processing aspects is supported by the Russian Science Foundation Grant No. 19-71-10092 (Sec.~\ref{Sec:QSC}). 
This work is supported by the Priority 2030 program at the National University of Science and Technology ``MISIS” under the project K1-2022-027 (analysis of quantum networks; Sec.~\ref{Sec:Hybrid}).

\bibliography{bibliography-rev.bib}

\end{document}